\documentclass[12pt]{article}
\usepackage{amsmath}
\usepackage{graphicx,epsfig}
\input epsf

\newcommand{\half}{\frac{1}{2}}
\newcommand{\nder}{\!\!\!\not\partial}

\begin{document}
\author{V.G. Ksenzov\footnote{State Scientific Center Institute for Theoretical and Experimental Physics, Moscow, 117218, Russia} \!\enskip and A.I. Romanov\footnote{National Research Nuclear University MEPhI, Moscow,115409, Russia}}
\title{Chiral condensate beyond one-loop approximation}
\date{}
\maketitle
\begin{abstract}

We study two-dimensional model with a fermion field and a self-interaction scalar field with the Yukawa interaction beyond the framework of the one-loop approximation. We show that in the model besides a phase transitions appear decays of the excited scalar states.

\end{abstract}

\section{Introduction}

In quantum field theory spontaneous symmetry breaking achieves by means of either scalar field which develops a non-vanish vacuum expectation value (v.e.v.) or dynamically produced. Dynamical breaking of a symmetry was formulated by Nambu and Jona-Lasinio (NJL) \cite{NJ}. In their paper analyzed a specific field model in four dimensions as a prototype for studying chiral symmetry breaking in interacting fermion systems. In recent years studies of those models have been extended in various way \cite{Bel} -- \cite{Braun}.

After NJL dynamical breaking was studied by Gross and Neveu (GN) in two-dimensional spacetime in the limit of a large number of fermion flavours $N$ \cite{GN}. These two models are similar but in contrast to the NJL model, the GN model is a renormalized and asymptotically free theory. The relative simplicity of both models is a consequence of the quartic fermion interaction.

In our previous papers \cite{KR} we investigated system of self-interaction massive scalar and massless fermion field with the Yukawa interaction in $(1+1)$-dimensional spacetime. In the limit of a large mass of the scalar field the model becomes equivalent to the GN model. The results were obtained by the functional integration using method of the saddle point. We have found a phase transition in the model in the one-loop approximation. It was shown that the chiral condensate vanishes and the symmetry is unbroken if $\cfrac{\lambda}{4Ng^2}$ is $O(1)$, where $\lambda$ and $g$ are coupling constants of scalar and fermion fields respectively.

In this paper we study the chiral condensate beyond the framework of the one-loop approximation. We find that in contrast to the one-loop approximation excited scalar state begins to decay with the growth of the effective energy of system if the coupling constants don't obey critical relation.

\section{Effective potential and chiral condensate}

In this paper we study a model with the Lagrangian density given by
\begin{equation}\label{eq1}
  L=L_b+L_f=\half(\partial_{\mu}\phi)^2-U(\phi)+i\bar{\psi}^a\nder\psi^a-g\phi(x)\bar{\psi}^a\psi^a,
\end{equation}
where $\phi(x)$ is a real scalar field, $\psi^a$ is a massless fermion field and the index $a$ runs from $1$ to $N$. The potential of the scalar field is
\begin{equation}\label{eq2}
  U(\phi)=\half m^2\phi^2+\frac{\lambda\phi^4}{24}.
\end{equation}

As it has already been mentioned above, the model is equivalent to the GN model in the limit of large $m$.

The Lagrangian (\ref{eq1}) is invariant under the discrete transformation
\begin{equation}\label{eq3}
  \psi^a\to\gamma_5\psi^a,\;\bar{\psi}^a\to-\bar{\psi}^a\gamma_5,\;\phi\to-\phi.
\end{equation}
The symmetry (\ref{eq3}) is broken either by the chiral condensate or by the scalar field which develops a non-vanishing vacuum expectation value. As it was shown in \cite{KR}, in this model the chiral condensate is connected with v.e.v. of scalar field.

We formally define the chiral condensate using the functional integral
\begin{equation}\label{eq4}
  \left\langle0|g\bar{\psi}^a\psi^a|0\right\rangle=\frac{1}{Z}\int D\phi D\bar{\psi}^aD\psi^ag\bar{\psi}^a\psi^a \exp\left(i\int d^2xL(x)\right),
\end{equation}
where $Z$ is a normalization constant. Eq. (\ref{eq4}) is rewritten as
\begin{equation}\label{eq5}
  \left\langle0|g\bar{\psi}^a\psi^a|0\right\rangle=\frac{1}{Z}\int D\phi e^{i\int d^2xL_b(x)}i\frac{\delta}{\delta\phi}\int D\bar{\psi}^aD\psi^a \exp\left(i\int d^2xL_f(x)\right).
\end{equation}
The fermionic Lagrangian is quadratic in the fields and we can integrate over them, getting
$$\left\langle0|g\bar{\psi}^a\psi^a|0\right\rangle=\frac{1}{Z}\int D\phi \frac{g^2 N\phi}{2\pi}\ln\frac{g^2\phi^2}{\Lambda^2}\times$$
\begin{equation}\label{eq6}
  \times\exp\left(i\int d^2x\left(L_b(x)-\frac{Ng^2\phi^2}{4\pi}\left(\ln\frac{g^2\phi^2}{\Lambda^2}-1\right)\right)\right),
\end{equation}
where $\Lambda$ is the ultraviolet cutoff. In order to integrate over the scalar field we decompose $\phi=\phi_0+\varphi$, where $\phi_0$ satisfies the classical equation of motion while $\varphi$ describes small fluctuations around the classical background. Then we take into account the quadratic terms of the scalar fluctuations $\varphi$, getting
$$S=S_{cl}+S_{qu},$$
where
\begin{equation}\label{eq7}
  \begin{split}
&S_{cl}=\int d^2x\half(\partial_{\mu}\phi_0)^2-\frac{m^2\phi_0^2}{2}-\frac{\lambda\phi_0^4}{24}-\frac{g^2N\phi_0^2} {4\pi}\left(\ln\frac{g^2\phi_0^2}{\Lambda^2}-1\right)\\
&S_{qu}=\int d^2x\half(\partial_{\mu}\varphi)^2-\frac{\varphi^2}{2}\left(m^2+\frac{g^2N}{2\pi}\left(\ln\frac{g^2\phi^2} {\Lambda^2}+2\right)+\frac{\lambda}{2}\phi_0^2\right).
  \end{split}
\end{equation}
The factor in front of the exponent in Eq.~(\ref{eq6}) is fixed at the point $\phi_0$, and integrating over $\varphi$, we obtain
\begin{equation}\label{eq8}
  \left\langle0|g\bar{\psi}^a\psi^a|0\right\rangle=\frac{Ng^2\phi_0}{2\pi}\ln\left.\frac{g^2\phi_0^2} {\Lambda^2}\right|_{\phi_0=\phi_m}.
\end{equation}
The quantity $\left\langle g\bar{\psi}^a\psi^a\right\rangle$ is a chiral condensate if only $\phi_m$ is a ground state value of bosonic field which is determined by equations
\begin{equation}\label{eq9}
  \left.\frac{dU_{\text{eff}}}{d\phi_0}\right|_{\phi_0=\phi_m}=0,\; \left.\frac{d^2U_{\text{eff}}}{d\phi_0^2}\right|_{\phi_0=\phi_m}>0,
\end{equation}
where
\begin{equation}\label{eq10}
  U_{\text{eff}}(\phi_0)=U(\phi_0)+\frac{g^2N\phi_0^2}{4\pi}\left(\ln\frac{g^2\phi_0^2}{\Lambda^2}-1\right)- \frac{f(\phi_0)}{8\pi}\left(\ln\frac{f(\phi_0)}{\Lambda^2}-1\right),
\end{equation}
and
\begin{equation}\label{eq11}
  f(\phi_0)=m^2+\frac{\lambda\phi_0^2}{2}+\frac{g^2N}{2\pi}\left(\ln\frac{g^2\phi_0^2}{\Lambda^2}+2\right).
\end{equation}
The effective potential $U_{\text{eff}}$ requires renormalization. Therefore we renormalize it, following Coleman and Weinberg \cite{CW} and Gross and Neveu \cite{GN} by demanding that
\begin{equation}\label{eq12}
  \left.\frac{d^2U_{\text{eff}}}{d\phi_0^2}\right|_{\phi_0=\kappa}=m_R^2.
\end{equation}
It is worth to note that in a $(1+1)$-dimensional spacetime there is only one $Z$-factor that appears in the renormalization of the scalar field mass, but the coupling constants obtain only finite contributions.

Then the renormalized chiral condensate is written as
\begin{equation}\label{eq13}
  \left\langle0|g\bar{\psi}^a\psi^a|0\right\rangle_R=\frac{Ng^2\phi_0}{2\pi}\left.\ln\frac{\phi_0^2}{\kappa^2} \right|_{\phi_0=\phi_{mR}},
\end{equation}
where $\phi_{mR}$ is determined by equations (\ref{eq9}), but for renormalized effective potential $U^R_{\text{eff}}$.
One can see that the renormalized chiral condensate vanishes if $\phi_{mR}^2=0$ or $\phi_{mR}^2=\kappa^2$. In the latter case the discrete symmetry (\ref{eq3}) is not restored, and the fermion field obtains a mass $\sim g\kappa$.

It should be noted that the third term in Eq.~(\ref{eq10}) describes the fermion multiloop contributions. This term gives an effect which does not appear within the one-loop approximation. Really, after renormalizating the value $f(\phi_0)$ must have the term $\sim\cfrac{g^2N}{2\pi}\ln\cfrac{\phi_0^2}{\kappa^2}$ which may be large negative value. Then $f(\phi_0)$ becomes negative value too, and effective potential $U_{\text{eff}}^R(\phi_0)$ obtains an imaginary part $\sim f(\phi_0)$. In such a case some excited scalar states having an energy $U_{\text{eff}}^R(\phi_0)>U_{\text{eff}}^R(\bar{\phi_0})$ must decay. Here $\bar{\phi_0}$ is a solution of the equation $f(\bar{\phi_0})=0$. The ground scalar states are stable due to $\phi_{mR}>\bar{\phi_0}$ as we shall see below. Our qualitative analysis we have verified by means of the numerical calculations.

\section{Obtained results}

\begin{figure}
\centering
\includegraphics[width=\linewidth]{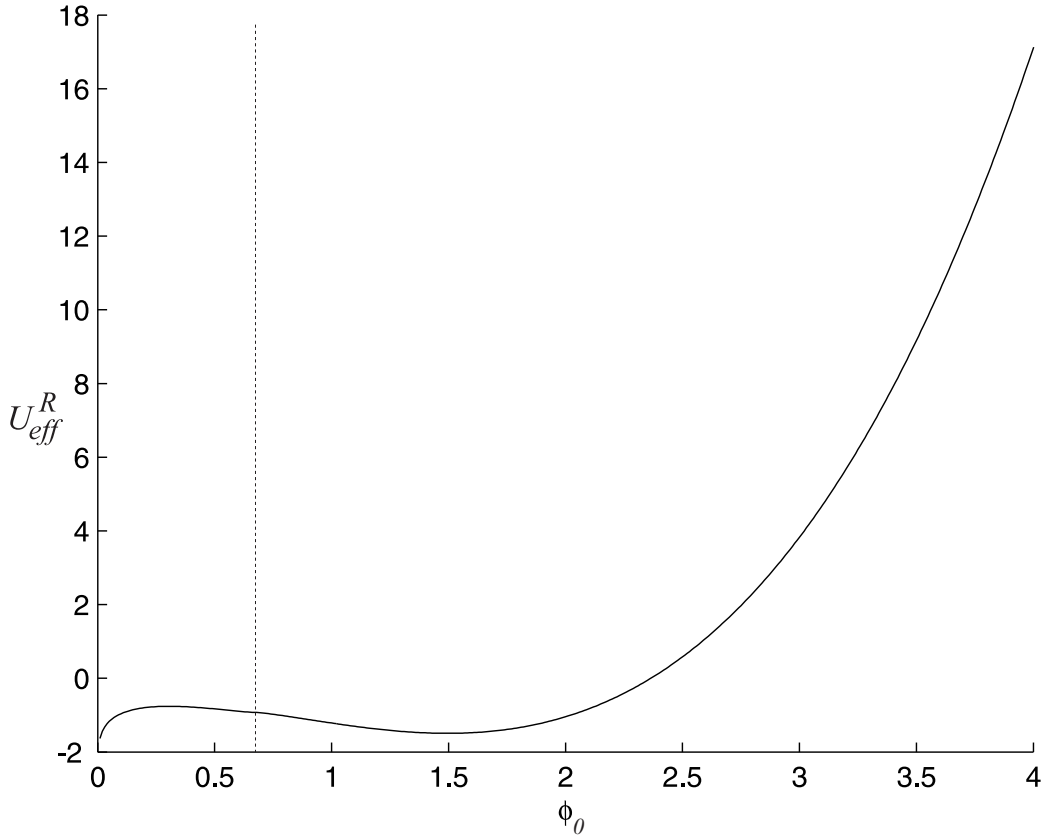}
\caption{The real part of effective potential for $g_0=0.4$, $\lambda_0=0.1$, $\kappa^2=10$. At the left from dotted line potential has an imaginary part.}
\label{Fig1}
\end{figure}

\begin{figure}
\centering
\includegraphics[width=\linewidth]{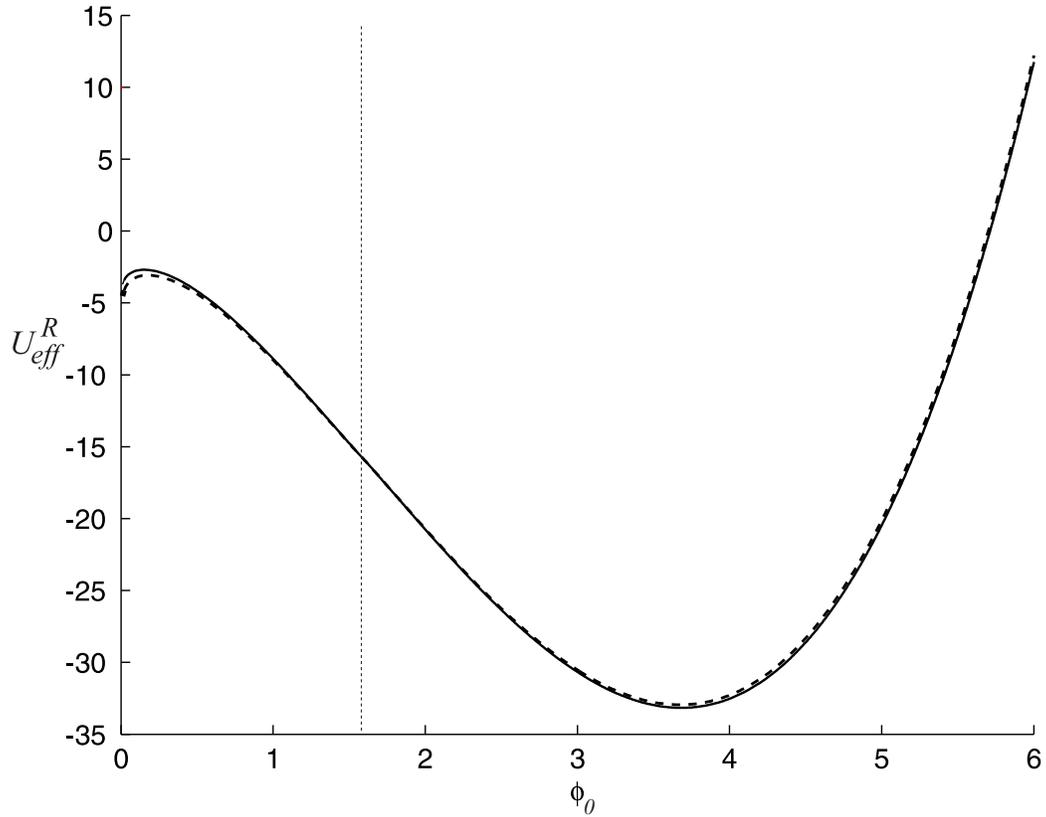}
\caption{The real part of effective potential for $g_0=0.66$, $\lambda_0=0.1$, $\kappa^2=10$ and two different values of ultraviolet cutoff: $\Lambda^2=150\text{ MeV}^2$ (solid line), $\Lambda^2=250\text{ MeV}^2$ (dashed line). At the left from dotted line potential has an imaginary part.}
\label{Fig2}
\end{figure}

By numerical solution of Eq.~(\ref{eq12}) we have obtained the renormalized effective potential $U_{\text{eff}}^R(\phi_0)$ and the chiral condensate $\left\langle0|g\bar{\psi}^a\psi^a|0\right\rangle_R$. The results of a numerical studies of effective potential are presented in Figs.~\ref{Fig1}-\ref{Fig2} as a function of the scalar field $\phi_0$. The parameters entering in Eq.~(\ref{eq12}) take the next values: $m_R=3\text{ MeV}$, $N=5$, $\lambda_0=\lambda/m_R^2$, $\kappa^2=10$. The values of parameters were arbitrarily chosen as an example only.

We investigated the behavior of the $U_{\text{eff}}^R(\phi_0)$ for different values of the coupling constant $g_0=g/m$. The effective potential flabbily depends on the coupling constant $\lambda_0$. Therefore we don't give it as a function of $\lambda_0$. Besides, we have checked (Fig.~\ref{Fig2}), that our results do not depend on the ultraviolet cutoff.

\begin{figure}
\centering
\includegraphics[width=\linewidth]{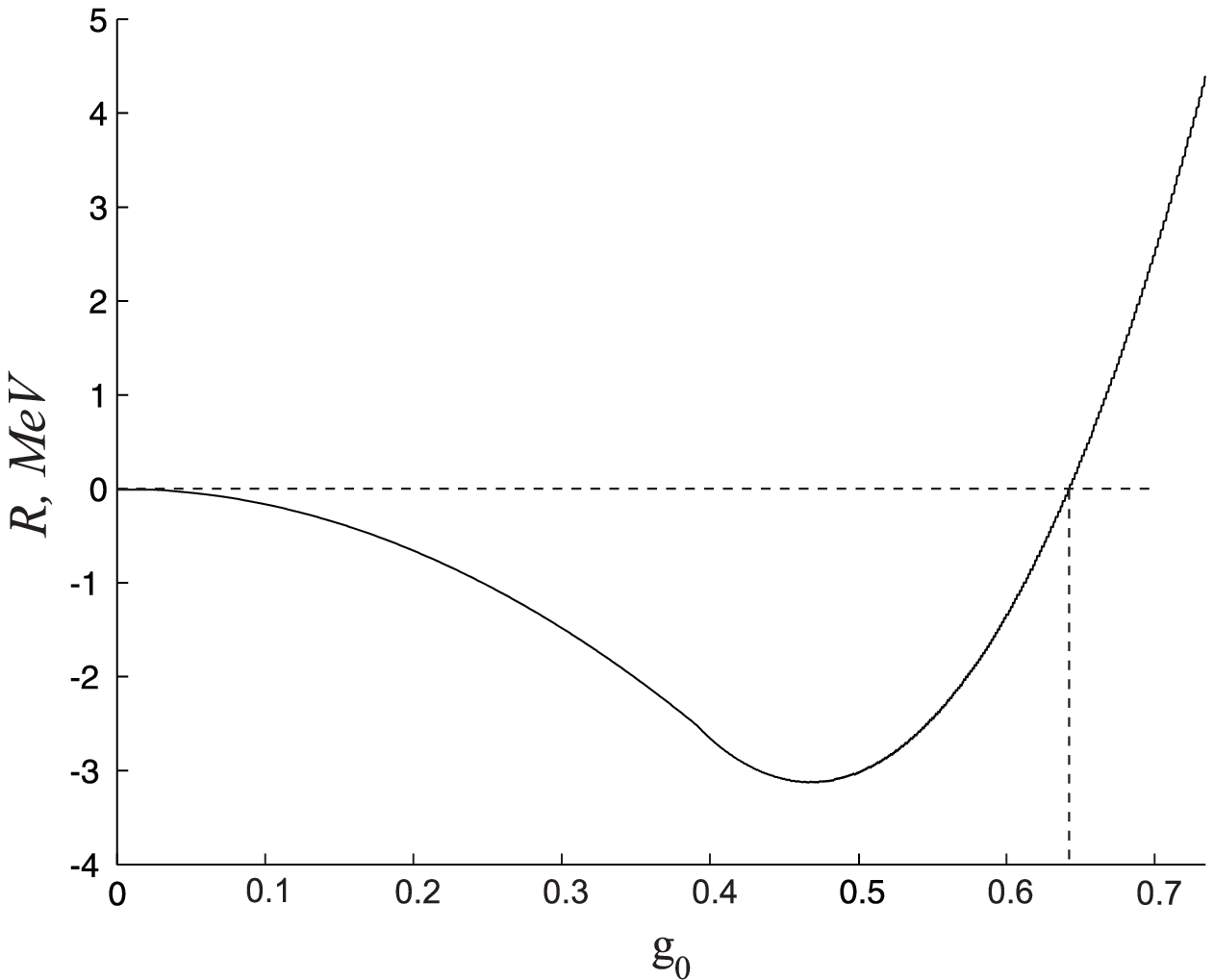}
\caption{The chiral condensate $R=\langle0|g_0\bar{\psi^a}\psi^a|0\rangle_R$ as a function of coupling constant $g_0$ for $\lambda_0=0.1$, $\kappa^2=10$.}
\label{Fig3}
\end{figure}

In Figs.~\ref{Fig1}-\ref{Fig2} dotted line separates the region where effective potential $U_{\text{eff}}^R(\phi_0)$ has an imaginary part (at the left) from the region where there is not imaginary part. The position of this dotted line strongly depends on the coupling constant $g_0$. For the small $g_0$ the region of complex effective potential practically disappears (see Eq.~(\ref{eq11}) and Fig.~\ref{Fig1}).


The ground state value $\phi_m$ is always more than the value of the scalar field $\bar{\phi_0}$ at which the imaginary part of effective potential disappears. Therefore the only excited states with the energy $U_{\text{eff}}^R(\phi_0)>U_{\text{eff}}^R(\bar{\phi_0})$ decay.

The chiral condensate is presented in Fig.~\ref{Fig3} as a function of the coupling constant $g_0$. We can see that the chiral condensate is equal to zero if $g_0\ll 1$ and $g_0\approx 0.65$. In the case of small $g_0$ the phase transition appears as it takes place in the one-loop approximation.
In the case $g_0\approx 0.65$ the chiral condensate vanishes. It corresponds to $\phi_m^2=\kappa^2$ (see Eq.~(\ref{eq13})), but the discrete symmetry (\ref{eq3}) is broken.

\section{Conclusions}

In this paper we analyzed the dynamical symmetry breaking in model consisting of a fermion fields and a self-interacting scalar fields with the Yukawa interaction beyond the framework of the one-loop approximation. We conclude that the model have a phase transition, as in the case of the one-loop approximation. Besides, we showed, that the excited scalar states begin decay with the growth of the effective energy of system at the coupling constants don't obey critical relation. Here we recall that the phase transition appears if $\lambda/4Ng$ is $O(1)$ \cite{KR}. This phenomenon is a result of the multiloop contributions. In the model the chiral condensate vanishes at two different coupling constants $g_0$. At the first of them there is a phase transition, at the second there is not. In the latter case the discrete symmetry is not restored.

We believe that the dynamical symmetry breaking and decay of the excited scalar states as exhibited by this model can help to analyze similar phenomena in more realistic models.

\begin{center}
Acknowledgments
\end{center}

We are grateful to K.G. Boreskov, O.V. Kancheli and V.A. Lensky for useful discussions.

\end{document}